\documentclass[twocolumn,aps,a4paper,superscriptaddress,prl,reprint]{revtex4-1}
\usepackage{graphicx}
\usepackage{subfigure}

\usepackage{latexsym}   
\usepackage{amsmath,amssymb,amsthm}

\usepackage{color}

\newcommand{\lgr}{\textcolor{navy}}
\newcommand{\gco}{\textcolor{purple}}

\definecolor{green}{rgb}{0.0, 0.666, 0.0}

\definecolor{darkgreen}{rgb}{0.55, 0.71, 0.0}

\definecolor{lightviolet}{rgb}{0.79, 0.28, 0.96}

\definecolor{navy}{rgb}{0.05, 0.40, 0.8}

\definecolor{purple}{rgb}{0.62, 0.0, 0.666}

\begin{document}

\title{Comment on ``Solvent-Induced Negative Energetic Elasticity in a Lattice Polymer Chain''}

\author{\firstname{L.} K. R. \surname{Duarte}} %

\author{\firstname{L.} G. \surname{Rizzi}} %

\noindent
{\bf Comment on ``Solvent-Induced Negative Energetic Elasticity in a Lattice Polymer Chain''}

	In a recent Letter, Shirai\,\&\,Sakumichi~\cite{shirai2023prl} presented a study focusing on 
the origin of a temperature-dependent negative contribution $G_U(T)$ to the elastic modulus $G(T)$ of 
hydrogels~\cite{yoshikawa2021prx}.~The authors support their findings through an energy-related 
stiffness $k_U(r,T)$ obtained from a single chain, with $r$ being the end-to-end distance of a random
walk on a 3D lattice.
	It is argued that the parameter $\varepsilon$ related to polymer-solvent interactions
is positive, so the energy $E_s$ of an elongated state should be smaller than the energy $E_b$ of 
a more compact state.
	We believe that the analogy between $G_U(T)$ and $k_U(r,T)$ might have misled their claim that
$G_U(T)<0$ when $\varepsilon>0$.

	It seems that the authors wanted to explore an approach similar to the one considered 
to compute the elastic moduli~\cite{yoshikawa2021prx}.
	Therefore, one might infer that the Maxwell relation used to obtain the $S$ contribution to
the stiffness, $k_S(r,T)=T \left(\partial k(r,T) / \partial T \right)_r$, is related to the 
free energy $A(r,T)$ and is given by $(\partial S/\partial r)_T = - (\partial f/\partial T)_r$, with 
the force evaluated as $f(r,T) = (\partial A(r,T)/\partial r)_T$ and $k(r,T)=(\partial f(r,T)/\partial r)_T$.
	Presumably, an analogy can be made between $f(r,T)$ and the stress $\sigma(\gamma,T)$, so the 
relation $(\partial S/\partial \gamma)_T = - (\partial \sigma/\partial T)_\gamma$ would then be used 
to compute the differential modulus~\cite{duarte2024softmatter} 
$K(\gamma,T)=(\partial \sigma(\gamma,T)/\partial \gamma)_T$, and its $S$ and $U$ contributions, that is, 
$K_S(\gamma,T)=T \left( \partial K(\gamma,T) / \partial T \right)_{\gamma}$ and 
$K_U(\gamma,T)=K(\gamma,T) - K_S(\gamma,T)$, for a given strain $\gamma$.
	Here, it is worth emphasizing that the elastic moduli are obtained only for small deformations, 
that is, $G(T)=\lim_{\gamma \rightarrow 0}K(\gamma,T)$, $G_S(T)=\lim_{\gamma \rightarrow 0}K_S(\gamma,T)$, 
and $G_U(T)=\lim_{\gamma \rightarrow 0}K_U(\gamma,T)$.
	So, rigorously, $k(r,T)$, $k_S(r,T)$, and $k_U(r,T)$ are analogous to the differential moduli and 
not to the elastic moduli.
	Oddly,  in Ref.~\cite{shirai2023prl}, the results for $k_U(r,T)$  were presented for arbitrary values of $r$ (see also Fig.~S1).

	These considerations revealed a subtle caveat: the experimental data of Ref.~\cite{yoshikawa2021prx} 
were analysed under a constant strain condition, while the results in Ref.~\cite{shirai2023prl} were obtained 
only for fixed values for $r$.
	As discussed in Ref.~\cite{duarte2023epje}, one must consider that $r$, just like the end-to-end 
distance of the chains in the gel network in the absence of external forces,  $r_0^{\,}=r_0^{\,}(T)$, is a 
temperature-dependent function, {\it i.e.}, $r(\gamma,T)=(1+\gamma)r_0^{\,}(T)$.
	Hence, one has that $dr =(\partial r/\partial T)_{\gamma} dT + (\partial r/\partial \gamma)_{T} d\gamma $, and the relation obtained from $dA = -S(r,T) dT + f(r,T)dr$ should be effectively replaced by 
$\left(  \partial S/ \partial r \right)_T = - \left(  \partial f/\partial T \right)_{\gamma} + \left(  \partial f/ \partial r \right)_T \left(  \partial r /  \partial T \right)_{\gamma}$, with the definition 
$\gamma = (r-r_0^{\,})/r_0^{\,}$.
	In practice, the $S$ contribution could be evaluated as $k_S(r,T)= \left. k_S(\gamma,T)\right|_{\gamma=(r-r_{0}^{\,})/r_{0}^{\,}}$, with $k_S(\gamma,T)= T \left(\partial k(\gamma,T) / \partial T \right)_{\gamma}$.
	If that was the case, in order to compare their results to the experimentally obtained elastic moduli of Ref.~\cite{yoshikawa2021prx}, they should have taken the limit where $\gamma \rightarrow 0$, {\it i.e.},
 $g(T)=\lim_{r \rightarrow r_{0}^{\,}(T)}k(r,T)$, $g_S(T)=\lim_{r \rightarrow r_{0}^{\,}(T)}k_S(r,T)$, and $g_U^{\,}(T)=\lim_{r \rightarrow r_{0}^{\,}(T)}k_U(r,T)$.

\begin{figure}[!ht]
\centering
\includegraphics[width=0.48\textwidth]{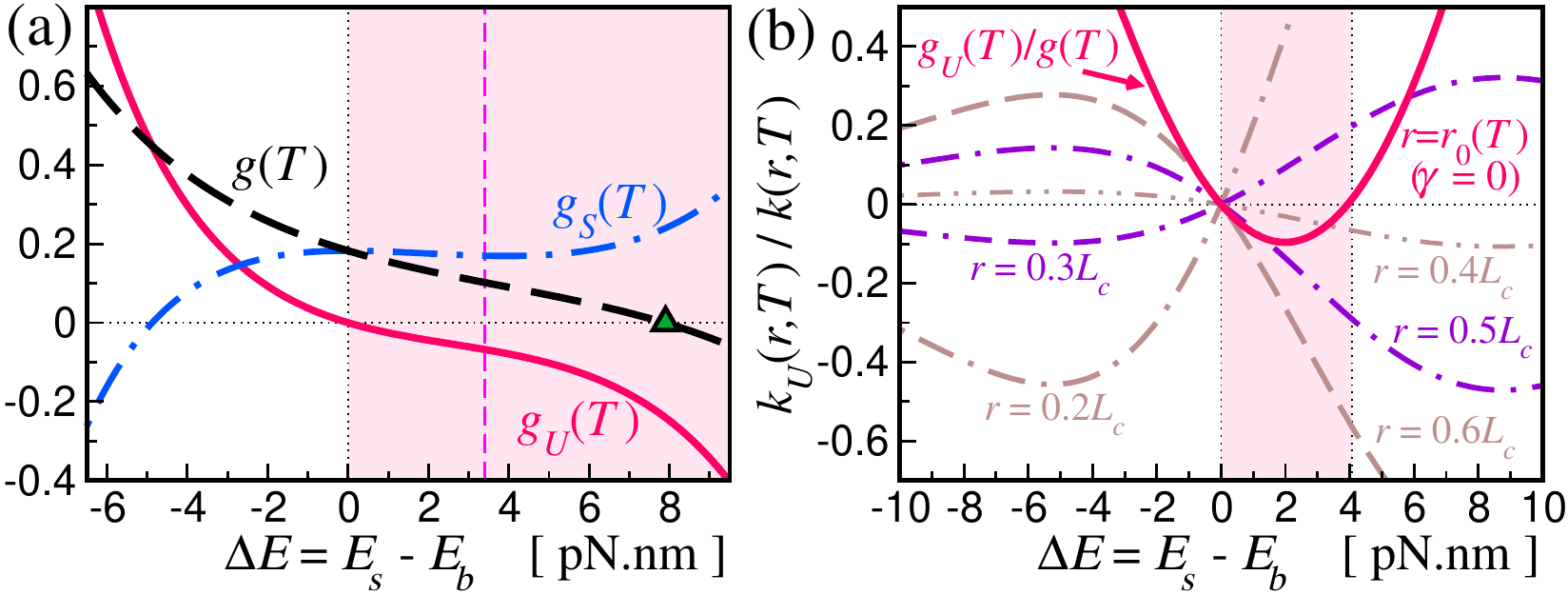}
\caption{Shaded (pink) regions denote $\Delta E$ where $g_U^{\,}(T)<0$. (a) $g(T)$, $g_S^{\,}(T)$, and $g_U^{\,}(T)$ for $\ell_s/\ell_b=-7.25$. 
	Vertical dashed line denotes $\Delta E=3.4\,$pN.nm used in Ref.~\cite{duarte2023epje} to describe the experimental data~\cite{yoshikawa2021prx}; green triangle indicates $\Delta E=k_BT\ln(-\ell_s/\ell_b)$ where $g(T)=0$.
(b)~Ratio $k_U^{\,}(r,T)/k(r,T)$ evaluated for arbitrary $r$ ({\it i.e.}, fixed fractions of the contour length $L_c$), and for $r=r_{0}^{\,}(T)$ ({\it i.e.}, $\gamma=0$), with $\ell_s/\ell_b=+7.25$.
	Results obtained at $T=288\,$K with
$k(\gamma,T)=g(T)/[1+\gamma/\gamma_b][1-\gamma/\gamma_s]$,
where $g(T)=k_BT[1+\exp(\Delta E/k_BT)]/\gamma_s(\ell_s-\ell_b)$,
$\gamma_s=(\ell_s-\ell_b)/[\ell_b+\ell_s \exp(-\Delta E/k_BT)]$,
$\gamma_b=\gamma_s \exp(-\Delta E/k_BT)$,
and
$\gamma = \gamma(r)$
(see Ref.~\cite{duarte2023epje} for details).
}
\label{fig:stiffness}
\end{figure}

	To illustrate our point, we consider the model described in Ref.~\cite{duarte2023epje}, which yielded quantitative agreement to the experimental data of Ref.~\cite{yoshikawa2021prx}.
	Figure~\ref{fig:stiffness}(a) shows that, if the correct analysis is considered, one finds that $g_U^{\,}(T)<0$ only when $\Delta E>0$, {\it i.e.,} $E_s > E_b$, and also that there is a $\Delta E$ where $g(T)=0$.
	Additionally, Fig.~\ref{fig:stiffness}(b) suggests that, by assuming arbitrary $r$ ({\it e.g.}, $r=0.3L_c$), one might incorrectly conclude that $k_U(r,T)<0$ only if $\Delta E<0$, {\it i.e.}, $E_s<E_b$, which is associated to $\varepsilon>0$.
	Our results indicate that, when $r=r_{0}^{\,}(T)$, {\it i.e.}, $\gamma=0$, one will find $g_U(T)<0$ 
(and, consequently, $G_U(T)<0$), only if $\Delta E>0$, that is, $\varepsilon<0$, in agreement to Eq.~29 in 
Ref.~\cite{duarte2023epje}, for any set of parameters.
	We note that, in contrast to the ``self-repulsive condition'' of Ref.~\cite{shirai2023prl},
most of previous approaches ({\it e.g.}, its Refs.~[18-22]), assume that $E_s>E_b$ ({\it i.e.}, $\varepsilon<0$), 
and those might agree with the phenomenology presented here.

	We thank the support received from FAPEMIG (Process APQ-02783-18) and CNPq (Grant 312999/2021-6).

~~

\noindent
L. K. R. Duarte$^{1,2}$ and L. G. Rizzi$^{1}$


\noindent
{\small 1.\,Departamento de F\'isica, Universidade Federal de Vi\c{c}osa (UFV), Av. P. H. Rolfs, s/n, , 36570-900, Vi\c{c}osa, Brazil.}

\noindent
{\small 2.\,Instituto Federal de Educa\c{c}\~ao, Ci\^encia e Tecnologia de Minas Gerais, P\c{c}.\,J.\,E.\,Dias, 87, 35430-034, Pte.\,Nova, Brazil.}


\pagebreak
\widetext

\fontfamily{qhv}\selectfont

\small

\begin{center}
We include below all the correspondence with PRL's editor (Manuscript ID: LTK1130), so one can draw its own conclusions.
\end{center}

\vspace{1.0cm}

\noindent
{\normalsize = = = = = = = = = = = = = = = =  [iii] Response to the Author's Reply (October 30, 2024)  = = = = = = = = = = = = = = =}

\vspace{0.5cm}

\noindent
Dear Editor,

\noindent
In our view, the authors tried to avoid our main criticism which is that they should have considered a constant-strain condition to obtain the thermodynamic derivatives instead of the constant $r$ condition, since the fundamental issue is to compare their results with the experimental results reported by [Yoshikawa et al., Phys. Rev. X 11 (2020) 011045].

\noindent
It is worth emphasizing that the main argument presented in our Comment is based on thermodynamic considerations and not on any specific model. In addition, by criticizing their analysis, we pointed out that it may lead to the uncertain conclusion that the origin of the negative contribution $G_U(T)$ to the elastic modulus is due to the repulsive character of the effective polymer-solvent interaction.

\noindent
We expected their Reply to address our primary criticism with pragmatics arguments and, eventually, results obtained from their model. Instead, the authors focused on trying to discredit our model, which we have introduced primarily to illustrate the misleading interpretations that can arise when the constant-strain condition is not properly considered.

\noindent
In any case, some statements in their Reply require clarification. For example, the authors claim that one of our assumptions is that ``(A) the distance between crosslinks of polymer networks in hydrogels ($r$) should be the end-to-end distance of the subchain with free ends, $r_0$.'' However, we emphasize that our definition of $r_0$ is not quite like that. As discussed in Ref.~[4] of our Comment, $r_0$ refers to the mean end-to-end distance of the (sub)chains inside the gel network in its as-prepared state. To make that clearer, we have revised the relevant sentence in our Comment's third paragraph to read: ``just like the end-to-end distance of the chains in the gel network in the absence of external forces.'' Interestingly, the schematic drawings in Fig.~1(b) and (c) in their Reply are nearly identifcal to those in Fig.~1 of Ref.~[4] in the Comment, where we present our model.

\noindent
In addition, in the attempt to invalidate our model, the authors state that such a definition (A) and the conclusion denoted by (C) should be unfounded since ``(D) the average distance between crosslinks of star-polymer gels at the as-prepared state ($r$) is a decreasing function of polymer mass concentration.'' While the statement (D) is completely fine {\it per se}, it does not contradict our definition of $r_0$ or any other arguments presented in our Comment, including those that lead to (C). In fact, as demonstrated in Ref.~[4] of our Comment, our model describes quantitatively all the experimental results reported by [Yoshikawa et al., Phys. Rev. X 11 (2020) 011045], which were intended to be explained by the authors in the Letter. It means that ou model can be used to describe not only the results related to the negative contribution $G_U(T)$ to the elastic modulus (which is the main point of the Letter) but also the whole set the experimental results involving gels with diferent molecular weight and concentration of precursor molecules, as well as netwrosk with different connectivities. Futhermore, as discussed in Ref.~[3] of our Comment [Duarte \& Rizzi, Soft Matter 20 (2024) 5616], our model sucessfully explains experimental observations in hydrogels based on star-polymers reported by [Kamata et al., Science 343 (2014) 873], where the increasing molar fraction of the hydrophobic precursor molecules in the network reduces the value of $r_0$, as expected. We acknowledge that the results presented in Refs.~[3,4] of our Comment highlight the effective character of our model, but that does not violate the statement (D) in any aspect.

\noindent
At the end of their Reply, the authors suggest that the comparison ``between the two models lacks physical significance'' due to ``differences in coarse-graining levels and spatial dimensions.'' However, the discussion on the character of the effective polymer-solvent interaction described by the parameter $\varepsilon$ is not based exclusively on our model. As we indicated in the Comment, it rather relies on the fact that in their self-avoiding walk-based model with $\varepsilon >0$, ``the energy $E_s$ of an elongated state should be smaller than the energy $E_b$ of a more compact state,'' and such a phenomenology can be compared and tested across different models.

\noindent
Finally, the authors state in the Reply that we claim that ``(B) in 'their' Letter, the interaction parameter $\varepsilon$ is positive, but it should be negative.'' In fact, this misrepresents our position. To be precise, we do not assert that models based on self-avoiding walks with positive $\varepsilon$ cannot yield negative contributions $g_U(T)$. Rather, we claim that the authors cannot give any definite answer and state that negative $g_U(T)$ occurs only when $\varepsilon$ is positive based solely on the results presented for $k_U(r,T)$ at arbitrary values of $r$ as in their original Letter. For a meaningful interpretation of their results in light of the experimental evidence, they should have computed the thermodynamic derivatives with the constant strain condition in the first place, as we argue in the Comment.

\vspace{2.0cm}

~

\pagebreak

\noindent
{\normalsize = = = = = = = = = = = = = = = = = = [iv] Reply to Referees (December 10, 2024) = = = = = = = = = = = = = = = = = =}

\vspace{0.5cm}

\noindent
Dear Editor,

\noindent
It is clear from the Referees' reports that they both agreed that our Comment is very helpful and scientifically pertinent as it unveils a substantial misinterpretation which is centrally related to the main conclusions of the original Letter by Shirai \& Sakumichi. Even so, the Referees were somewhat reluctant to recommend it for publication in PRL.

\noindent
As it happens, the Referees' reports presented clearly contrasting recommendations. While Referee A did not support the publication of our Comment on the argument that it is ``easy to arrive at this misinterpretation'', Referee B ``strongly recommended'' its publication, but elsewhere as a more detailed article.

\noindent
Hence, below we include a point-by-point discussion about their concerns and suggestions.

\vspace{0.5cm}

\noindent
Remarks from Referee A: \lgr{``The Comment and the response provide helpful context for the understanding of the Letter. While the Comment is based on a misinterpretation of the results (as explained very well in the response) it is nonetheless easy to arrive at this misinterpretation, and it is therefore useful. This might however not be sufficient to warrant publication of the Comment, given the stringent criteria.''}

\noindent
Reply:
Although the misinterpretation in the Letter was easily identified by the Referee, the fact that it was not spotted during the stringent review process of PRL indicates that it may not be the case even for some of its expert readers. Additionally, Referee B ``strongly recommended'' the publication of our Comment, though as a more detailed article to provide length to our argument. Even so, it might become clear to Referee B (as it seems to be to Referee A) that our Comment is very specific in a way that its content exclusively addresses the Letter's analysis and conclusions so that there are no other journals, forums, etc, where it could be published.

\noindent
And it is worth emphasizing that in their Reply, the authors of the Letter did not provide any evidence that they could refute our criticisms. Hopefully, the Referees might also take that into consideration when pondering the publication of our Comment in PRL.

\vspace{0.5cm}

\noindent
Remarks from Referee B: \lgr{``The manuscript critiques Shirai and Sakumichi's study regarding the temperature-dependent negative contribution to the elastic modulus of hydrogels. The authors highlight the importance of the temperature dependence of the unperturbed dimension of polymers. Specifically, they defined moduli as the differential moduli at the limit of the undeformed state. Such moduli exhibit negative values according to the analysis reported by the authors previously (Ref. 4).
The reviewer finds that the presented criticism is important and worth reporting somewhere. However, it is not suitable as a Comment in PRL. The reason is that the format seems incapable of accommodating the discussion. Indeed, the most critical issue, the temperature dependence of $g_U(T)$, is not described here. Without this function, the discussion in Fig. 1(a) cannot be followed. The explanation of Fig. 1(b) is not sufficiently presented and is hard to understand. The in-line equations are difficult to follow.
In short, I strongly recommend that the authors publish this issue as a new paper elsewhere with sufficient length for their argument.''}

\noindent
Reply:
We thank the positive opinion the Referee has on our Comment and acknowledge that he/she ``strongly recommends'' its publication, even though elsewhere. Conversely, Referee A argued that our criticism is not worth publishing in PRL because it is ``easy to arrive'' at our conclusions.

\noindent
As we argued in the Reply to Referee A, the criticism we present in our Comment is very specific, as it addresses the analysis and conclusions of this particular Letter, so we think that the arguments presented in our Comment would be devoid of purpose elsewhere. In addition, it became clear from the reports that the publication of our Comment in PRL would encompass the concerns of many of its interested readers (especially because it will allow the authors to present their Reply).

\noindent
Referee B also suggested that: (i) the ``critical'' discussion should be around ``the temperature dependence of $g_U(T)$;'' and that (ii) Fig.~1 in our Comment might not be so easy to understand. We acknowledge the Referee's remarks, but these two statements only reinforce the problems related to the presentation and analysis of the results in the original Letter and, ultimately, indicate how our Comment discloses these problems in a straightforward and understandable manner. The reason for this is that Fig.~1, in particular, was incorporated in our Comment chiefly to facilitate the analogy with the results and analysis presented in the Letter (one should note that even the colors are similar!). Also, the appropriate discussions and presentation of results in terms of $g_U(T)$ that Referee B may have wished for are already presented in Ref.~[4] of our Comment, where we include direct quantitative comparisons between theory and experiments (so the interested readers would benefit from this reference when reading the Comment as well). In any case, we must emphasize that, unlike the Referee's view, the ``most critical issue'' is how the authors in the original Letter did not consider the constant-strain condition to perform the correct thermodynamic derivatives to recover $g_U(T)$ in the proper limit of small strains. Their results are based not on $g_U(T)$ but instead on the so-called stiffness $k_U(r,T)$ evaluated at arbitrary values of the end-to-end distances $r$, and that hampered all their analyses. Consequently, these unwary considerations led them to the questionable conclusion that the effective polymer-solvent interaction should be repulsive for the negative contribution to the elastic modulus $G_U(T)$ to be observable as in the experiments reported by [Yoshikawa et al., Phys. Rev. X 11 (2020) 011045], which were performed under the constant-strain condition. That is the ``critical issue'' and the whole point of our Comment.

\noindent
In view of this, we think that not only the scientific context but also all the technical aspects necessary for the comprehension of our arguments were already included in the original version of the Comment (which not have been modified since), therefore we are convinced that this format is the most suitable way of presenting our criticisms to the original Letter by Shirai \& Sakumichi.



\pagebreak

\noindent
{\normalsize = = = = = = = = = = = = = = = = = [v] Referees' second report (December 24, 2024) = = = = = = = = = = = = = = = = =}

\vspace{0.5cm}

\noindent
Remarks from Referee A: \lgr{``The current communication seems to willfully misunderstand my remarks. I wrote `While the Comment is based on a misinterpretation of the results (as explained very well in the response)' which clearly means that I side with the response to the Comment, i.e. the fact that the comment misinterprets the results in the Letter. This is at odds with the statement that `Although the misinterpretation in the Letter was easily identified by the Referee.' I see no need to change my judgment and would now encourage the writers of the Comment to take the discussion elsewhere, as recommended by the other referee.''}

\vspace{0.5cm}

\noindent
Remarks from Referee B: \lgr{Although I am partly convinced by the authors, their revision does not meet my suggestions. Thus, I cannot change my rating; this manuscript does not fit the standard for PRL, although the study is worth publishing elsewhere.}

\vspace{19cm}

~

\pagebreak

\noindent
{\normalsize = = = = = = = = = = = = = = = = = = = = Formal appeal (December 29, 2024) = = = = = = = = = = = = = = = = = = = =}

\noindent
Dear Editor,

\noindent
Based on the Referees' second reports [v], we find that both are less reluctant about the content of our Comment [i] since they now encourage and recommend its publication, even though elsewhere.

\noindent
As it is argued in our response to the Referees' first reports [iv], our Comment is very specific as it centrally addresses the analysis and conclusions of the original Letter by Shirai \& Sakumichi, so its publication elsewhere would be purposeless.
Also, we are convinced that our Comment will be valuable to the PRL readers, especially because they will recognize that
an opportunity was given to the authors of the Letter to present their Reply.

\noindent
We acknowledge that we mistook the document called ``response'' by Referee A in the first report [iv] as our response to the formal Reply [iii], not the authors' Reply itself [ii]. Still, Referee A reaffirmed just that it is ``easy'' to identify that we have misconstrued the original Letter, but because no arguments were given for that, it sounded like an arbitrary (almost non-scientific) judgement. Hence, we insist that, in contrast to what is claimed by Referee A [v], our response to the formal Reply [iii] clearly indicates that the authors in their Reply [ii] did not provide any evidence that they could refute our criticisms. 

\noindent
Thus, what remained from the Referees' reports regard not the criticisms we have put forward, which are clearly expressed and contextualized in the Comment [i], but instead have to do with the format itself.

\noindent
Obviously, unless the Editor allows us to increase the length of the Comment (which we think is not the case), there might be no room for expanding the in-line equations or including new mathematical expressions, as requested by Referee B [iv]. Even so, we do not think that this is strictly needed since we already mentioned in the Comment [i] that an expression for $G_U(T)$ is provided in Ref. [4] (we also recall that [4] includes successful quantitative comparisons between our model's results and the experimental data [2] that the original Letter attempted to describe). We note that any additional expressions would be specific for the example we have included primarily to illustrate doubtful conclusions that may arise from the analysis the authors did in the Letter. However, it is worth clarifying that the absence of expressions like the one indicated by Referee B does not affect the main criticisms we present in the Comment [i], which are based instead on thermodynamic considerations and specifically address the misleading analysis performed in the original Letter through the so-called stiffness $k_U(r,T)$ (here we refer also to our responses to the Referees' reports [iv] and to the formal Reply [iii] for additional discussions).

\noindent
Hence, we want to make a formal appeal and request that our submission be considered by a member of the PRL's Editorial
Board. In particular, we would be grateful if our Comment is sent to\gco{(*)} Prof.~Daniel J. Read, Prof.~Xiaoming Mao, or Prof.~Friederike Schmid, from the Polymer Physics divisional area.

\noindent
For convenience, I have included below all the documents related to our submission, including:\\
i. our Comment (which has not been modified since its original submission);\\
ii. the formal Reply from the authors\gco{(**)};\\
iii. our response to their Reply;\\
iv. our response to the Referees' first reports;\\
v. the Referees' second reports.

\noindent
Eventually, we also ask you to kindly forward the current page of this document to the Editorial Board member as it
corresponds to our response to the Referees' second reports.

\vspace{5.0cm}

\noindent
\gco{(*) We clarify that the Divisional Associate Editor that gave the final report on our Formal Appeal was not one of them.}

\noindent
\gco{(**) For obvious reasons, we did not include the formal Reply of the authors of the Letter in this document, but some of its content can be inferred from our response [iii].}

\pagebreak

~

\vspace{0.5cm}

\noindent
{\normalsize = = = = = = = = = = = = = Final report of the Divisional Associate Editor (March 12, 2025)    = = = = = = = = = = = = = =}

\vspace{0.5cm}

\noindent
\lgr{``Thank you for your appeal regarding the decision on your Comment submission, 'Comment on Solvent-induced negative energetic elasticity in a lattice polymer chain.' I have carefully reviewed your appeal, along with the referee reports and the authors' formal Reply. After thorough consideration, I regret to inform you that the decision to reject your Comment stands.''}

\vspace{0.15cm}

\noindent
\lgr{``You argue that your Comment is highly specific to the analysis and conclusions of the original Letter, making publication elsewhere purposeless. However, Physical Review Letters maintains strict criteria for Comments: they must establish a central error in the original Letter that is clear and impactful. The referees have determined that your Comment does not meet this threshold, as it does not expose a fundamental flaw.''}

\vspace{0.15cm}

\noindent
\lgr{``While your Comment provides context for understanding the Letter, both referees have stated that it does not meet the PRL standard in terms of clarity and accessibility. A Comment should concisely and directly challenge key aspects of a Letter in a way that is immediately comprehensible to PRL's broad readership. The referees have judged that the structure and presentation of your Comment do not achieve this goal.''}

\vspace{0.15cm}

\noindent
\lgr{``You suggest that Referee A did not provide a scientific argument to support their statement that your Comment misinterprets the original Letter. However, Referee A has clearly stated that the authors' Reply successfully demonstrated that the Comment is based on a misinterpretation of the Letter's findings. The role of a Comment is to correct errors, not to highlight possible misinterpretations that can arise. The fact that such a misinterpretation is 'easy to arrive at,' as acknowledged by Referee A, does not provide sufficient grounds for publication in PRL.''}

\vspace{0.15cm}

\noindent
\lgr{``You argue that Referee B's concerns regarding missing expressions and technical clarity should not impact the validity of your Comment. However, clarity and accessibility are crucial publication criteria for PRL. 
The referees have indicated that your Comment does not sufficiently explain key aspects, such as the temperature dependence of $g_U(T)$, making it difficult to follow.''}

\vspace{0.15cm}

\noindent
\lgr{``Additionally, Referee B explicitly recommended that your work would be better suited for a longer article elsewhere where you can present the necessary details in full. While you note that some of these details are available in your previous works, a PRL Comment must be self-contained and fully comprehensible on its own. The suggestion to publish a longer version is not a dismissal of your argument but rather a recognition that the discussion requires more space than PRL can accommodate.''}

\vspace{0.15cm}

\noindent
\lgr{``Given the referees' consistent assessments and PRL's publication criteria, I must uphold the decision to decline publication of your Comment. However, I encourage you to consider submitting a more detailed version of your work to a journal where you can fully develop your analysis in a format that allows for the necessary depth and clarity.''}

\vspace{0.15cm}

\noindent
\lgr{``Thank you for your engagement in this scientific discussion, and I appreciate the time and effort you have dedicated to this process.''}


\begin{thebibliography}{99}

\bibitem{shirai2023prl} 
Shirai \& Sakumichi.
Phys. Rev. Lett. {\bf 130}, 148101 (2023).

\bibitem{yoshikawa2021prx} 
Y. Yoshikawa, N. Sakumichi, U. I. Chung, and T. Sakai. 
Phys. Rev. X {\bf 11}, 011045 (2021).

\bibitem{duarte2024softmatter}
Duarte \& Rizzi.
Soft Matter {\bf 20}, 5616 (2024).

\bibitem{duarte2023epje}  
Duarte \& Rizzi.
Eur. Phys. J. E {\bf 46}, 52 (2023).

\end{thebibliography}
\end{document}